# Governed Auditable Decisioning Under Uncertainty: Synthesis and Agentic Extension

Oleg Solozobov[1*]

[1] Independent Researcher (Global)

[*] Correspondence: Oleg Solozobov (dev404ai@gmail.com)

ORCID: https://orcid.org/0009-0009-0105-7459

## Abstract

When automated decision systems fail, formally compliant organizations often cannot reconstruct what happened or why. This paper synthesizes an operational governance evidence framework — structural accountability collapse diagnostics, decision trace schemas, evidence sufficiency measurement, and label-free monitoring — into an integrated chain and analytically assesses its transferability across four decision system architectures. The comparison reveals a governance coverage gradient: deterministic rule engines achieve near-full DES-property fillability (five of six properties by construction), hybrid ML+rules systems partial, classical ML only minimal, and agentic AI encounters structural breaks. We introduce the cascade of uncertainty, showing how governance failures propagate through serial dependencies between framework layers. For agentic systems, we identify three structural breaks — decision diffusion, evidence fragmentation, and responsibility ambiguity — and propose corresponding analytical extensions. Four propositions formalize the gradient, cascade compounding, delegation-depth effects, and extension sufficiency, establishing the framework's valid operating envelope.

## 1. Introduction

When automated decision systems fail, post-incident investigations repeatedly discover that formally compliant organizations cannot reconstruct the decision process from available evidence. The gap between formal compliance and operational accountability is well-documented: organizations that satisfy regulatory requirements on paper may nonetheless lack the evidentiary infrastructure to explain, after the fact, how a specific decision was reached and whether it was appropriate. Accountability requires not merely that rules exist but that decision processes are reconstructable from available governance artifacts (Novelli et al., 2024).

Four prior studies by the author developed an operational governance evidence framework to address this recurring pattern: the governance evidence framework (Solozobov, 2026b) diagnosed structural accountability collapse and its four modalities, the decision trace schema (Solozobov, 2026a) specified a Decision Event Schema for decision trace capture, the evidence sufficiency model (Solozobov, 2026c) defined evidence sufficiency metrics for environments with delayed ground truth labels, and the label-free monitoring layer (Solozobov, 2026e) completed the chain with detection of governance degradation without outcome data. Together, these contributions form what we term the N4 governance evidence chain (named for the four foundational papers) — a layered architecture progressing from diagnosis through evidence capture, sufficiency measurement, and continuous monitoring.

However, the generalizability of this framework remains untested. The algorithmic accountability literature has produced a rich set of normative frameworks, audit methodologies, and impact assessment protocols (Moss et al., 2021; Ada Lovelace Institute; AI Now Institute; Open Government Partnership, 2021; Mökander et al., 2021), but these operate at the level of organizational process or regulatory requirement rather than at the level of technical evidence architecture; they specify what should be evaluated but not how the underlying evidence is produced, validated, or maintained across system architectures. The N4 framework addresses this evidentiary substrate: it defines the minimum evidence requirements that any accountability mechanism presupposes.

This framework was developed and analytically assessed in the context of risk decision systems — specifically, financial fraud detection and regulatory compliance pipelines that combine machine learning models with business rules. The present paper asks three questions that extend beyond this origin domain. First, does the framework generalize across different decision system architectures? Automated decision-making in practice spans a wide spectrum, from deterministic rule-based systems with binary logic to machine learning systems with learned decision boundaries (Hackney & Huggins, 2023). If the framework addresses only one architecture type, its theoretical contribution is narrower than claimed. Second, how do governance failures at one framework layer affect other layers? The four components — diagnostic theory, trace schema, sufficiency measurement, label-free monitoring — were developed as modular contributions, but they may interact in ways that create compound vulnerabilities. Third, can the framework extend to agentic AI systems, where decisions are fragmented across multiple autonomous agents with delegated authority and emergent coordination?

The third question is particularly urgent. Agentic AI is rapidly proliferating across enterprise decision environments (Mukherjee & Chang, 2025), and AI decisions increasingly occur outside the window in which human intervention can reliably act (Driver, 2026). Agentic systems decompose goals into sub-tasks, delegate them, and produce composite outcomes through multi-step reasoning chains involving orchestrators, external APIs, and plugins whose diverse roles complicate decision-rights determination (Liu et al., 2024), creating what Mukherjee and Chang term an accountability vacuum — a void in which responsibility evaporates because agentic AI has its own operational autonomy (Mukherjee & Chang, 2025).

Agentic systems do not merely extend the challenges of classical ML or hybrid architectures: where prior architectures introduce bounded opacity that can be mitigated through explainability artifacts and oversight checkpoints, agentic systems introduce structural fragmentation of decisions, evidence, and responsibility. Existing governance frameworks — including those designed for high-risk AI under the EU AI Act — were developed for systems with identifiable decision points and attributable decision-makers (Kaminski, 2019), and agentic systems challenge these assumptions at an architectural level.

Framed in human-technology terms: accountability is a relation among designers, operators, auditors, vendors, affected subjects, and regulators, mediated by the evidence infrastructures these actors share. The framework asks when that mediation is technically reconstructable versus when it becomes structurally impossible. When agents decompose and delegate decisions across commercial API boundaries, the relations that would ordinarily support accountability — principal to agent, reviewer to decision, subject to explanation — lose their evidentiary substrate. The N4 chain specifies the minimum substrate any such relation presupposes; the structural breaks identified here locate where it dissolves.

This paper makes three contributions. First, it synthesizes the N4 framework into an integrated chain and analytically assesses its transferability across four architectures, revealing a governance coverage gradient: deterministic (near-full fillability), hybrid (partial), classical ML (minimal), agentic AI (structural breaks, zero fillable). Second, it introduces the cascade of uncertainty — a structural property by which governance failures at one layer propagate downward, mapping the four N4 components into a serial dependency chain with compound failure modes that no single component detects. Third, it identifies three structural breaks agentic AI introduces — decision diffusion, evidence fragmentation, and responsibility ambiguity — and proposes analytical extensions preserving the governance properties of the prior studies (Solozobov, 2026b; Solozobov, 2026a; Solozobov, 2026c; Solozobov, 2026e).

The analysis is grounded in a purpose-built governance evaluation framework that formalizes the comparison as a structured evaluation of six DES properties across four architectures, producing feasibility matrices that quantify coverage as the proportion of properties that are fillable, partially fillable, unfillable, or opaque. Four propositions formalize the gradient, cascade compounding, delegation-depth effects, and extension sufficiency: P1 is an analytical derivation; P2 is empirically testable through instrumented pipelines; P3 and P4 are analytical expectations derived from the framework's structure, establishing a research agenda for empirical validation.

The paper proceeds as follows. Section 2 synthesizes the four prior papers into an integrated framework. Section 3 applies the framework to four system architectures using the governance evaluation framework, establishing the governance coverage gradient. Section 4 introduces the cascade of uncertainty and maps it to the framework's components. Section 5 examines the agentic structural breaks and proposes framework extensions. Section 6 defines boundary conditions and limitations. Section 7 discusses implications and future work.

# 2. The Operational Governance Evidence Framework

The operational governance evidence framework developed across four prior studies constitutes an integrated analytical chain that addresses a recurring failure pattern in automated decision systems: organizations that satisfy formal compliance requirements nevertheless cannot reconstruct the decision process when failures occur. Each paper contributes a distinct layer to this chain, and together they provide an integrated governance evidence infrastructure from diagnostic theory through operational measurement. This section synthesizes the four contributions into an integrated model that subsequent sections will test against different system architectures.

## 2.1. Structural Accountability Collapse and Four Modalities

The governance evidence framework (Solozobov, 2026b) identifies the root phenomenon: structural accountability collapse (SAC), defined as the condition in which architectural properties of automated decision systems — velocity, scale, and opacity — exceed the design capacity of governance infrastructure, causing governance evidence to degrade while compliance artifacts persist. The analytical contribution is a distinction between compliance evidence (demonstrating that a process was followed) and governance evidence (enabling reconstruction of whether the decision-making process was adequate to the risks it addressed). Three jointly necessary properties define governance evidence: reconstructability, evaluability, and contemporaneity.

The paper establishes four modalities through which SAC manifests: evidence gaps (governance-relevant data is not captured), responsibility diffusion (accountability cannot be attributed to identifiable actors or roles), decision opacity (the decision logic cannot be reconstructed from available artifacts), and feedback failure (outcome signals do not reach the governance layer in time to support oversight). These modalities are structurally distinct — each produces a different type of governance evidence deficit, and each requires different remediation. The four modalities form the diagnostic foundation for the entire framework, providing a taxonomy that subsequent papers operationalize and measure (Cobbe et al., 2021).

## 2.2. Decision Trace Schema

The decision trace schema study (Solozobov, 2026a) operationalizes the governance-compliance distinction through a six-property Decision Event Schema (DES) that captures the provenance chain from input features through model inference to final decision. The schema defines the minimum evidence record that must be preserved for each decision event to enable post-hoc audit and reconstruction. As of DES v0.3.0, each event must declare an explicit *schema_version*, and the six property groups carry formal numbering: decision context (semantic completeness), decision logic (causal reconstructability), decision boundary (cross-system coupling), decision quality indicators (evidence of decision quality), human override record (independence of human judgment), and temporal metadata (temporal evidence and cryptographic integrity). Where compliance logging typically records that a decision was made and by which system, the DES captures the full decision context: what data was available, what model version processed it, what features were computed, what thresholds applied, and what the decision outcome was.

The decision trace concept finds independent support in recent accountability frameworks, which converge on provenance-based audit trails, chronological decision records, and integrated data trails that enable step-by-step reconstruction of AI system activity (Huynh et al., 2020; Anumula, 2022; Natta, 2025). This convergence confirms that the DES addresses a structural need rather than an idiosyncratic design choice.

The DES further specifies completeness criteria: a trace is governance-adequate only if it permits reconstruction of both the decision and the decision context. DES v0.3.0 operationalizes this through a tiered evidence model (*lightweight*, *sampled*, *full*) that defines progressive completeness requirements — lightweight traces capture required fields only, sampled traces add abbreviated logic and context, and full traces populate all six property groups completely. This distinction matters because many existing logging systems capture sufficient data to replay a decision (given the same model and inputs, produce the same output) but insufficient data to evaluate whether the decision process was appropriate (whether the right data was available, whether the model was fit for purpose, whether override conditions were met).

## 2.3. Evidence Sufficiency Under Delayed Labels

The evidence sufficiency study (Solozobov, 2026c) addresses a measurement problem that arises once decision traces exist: how to determine whether accumulated governance artifacts contain enough information to support accountability judgments. This problem is compounded by label delay — the condition in which ground truth outcomes (whether a decision was correct) arrive days, weeks, or months after the decision was made, or in some cases never arrive at all.

The paper defines evidence sufficiency metrics that evaluate governance artifact quality independently of outcome labels. Rather than asking whether decisions were correct (which requires labels), sufficiency measurement asks whether the evidence archive is complete enough to support reconstruction and evaluation if and when accountability is required. This reframing separates the governance question (can we reconstruct?) from the performance question (was the decision right?) and enables continuous governance assessment even in domains with permanent label absence. Audit-ready logging architectures in clinical AI deployment illustrate this principle: logging systems that capture model versions, input data identifiers, and inference parameters enable reproducibility of AI determinations regardless of when or whether diagnostic ground truth becomes available (Joseph, 2023).

### 2.4. Label-Free Monitoring

The label-free monitoring study (Solozobov, 2026e) completes the framework with a monitoring layer that detects governance degradation without requiring outcome labels. Where the evidence sufficiency study (Solozobov, 2026c) measures the static sufficiency of governance artifacts at a point in time, the monitoring study addresses the dynamic problem: detecting when governance evidence quality is deteriorating over time.

The monitoring approach uses distribution shift detection and evidence quality signals as proxies for governance health. If the statistical properties of decision traces change (feature distributions shift, decision time patterns alter, override rates diverge from baselines), this may indicate that the governance evidence being collected no longer represents the actual decision process. Modern logging architectures that treat individual executions as immutable historical events support complete system state reconstruction for any historical time point, providing the foundation for detecting temporal drift in governance evidence quality (Nallapu, 2025).

Label-free monitoring is particularly important for systems where outcome feedback is structurally delayed or absent — precisely the systems where governance failures are most difficult to detect. By monitoring the evidence production process rather than decision outcomes, the framework provides continuous oversight capability that does not depend on external validation signals.

### 2.5. The Integrated Chain

The four contributions form a serial dependency chain: sufficiency (Solozobov, 2026c) cannot be measured without a trace schema (Solozobov, 2026a); label-free monitoring (Solozobov, 2026e) cannot detect degradation without sufficiency baselines; none of these operational layers function without the diagnostic theory (Solozobov, 2026b) defining what governance evidence failure looks like.

The chain was developed and analytically assessed in financial fraud detection and regulatory compliance systems. Component specifications that the present paper's cross-architecture scoring depends upon are documented in the open-access governance evaluation framework repository (Solozobov, 2026d). The remaining sections test the chain's generalization across architectures, analyze the cascade of uncertainty that connects governance layers, and identify the structural breaks that emerge in agentic AI.

## 3. Cross-Architecture Governance Comparison

The operational governance evidence framework was developed in the context of risk decision systems — specifically, financial fraud detection pipelines combining learned models with business rules. A natural question is whether the framework generalizes beyond this origin domain. This section applies the N4 governance criteria (reconstructability, evaluability, contemporaneity, and the four SAC modalities) to four progressively more complex decision system architectures: deterministic rule engines, classical ML with human oversight, hybrid ML+rules systems, and agentic AI. The comparison reveals a governance coverage gradient measured by DES-property fillability: deterministic systems achieve near-full fillability (five of six properties by construction), hybrid systems achieve partial fillability, classical ML systems achieve only minimal fillability, and agentic systems encounter structural breaks with zero fillable properties.

### 3.1. Deterministic Rule-Based Systems

Deterministic rule-based systems are the strongest fit for the N4 framework. Decisions follow explicit human-authored rules applied to structured inputs and can be reconstructed from rule set and input data alone: reconstructability is inherent (same inputs + rules -> same decision), evaluability follows from rule transparency, and contemporaneity requires only that inputs are logged at decision time. Such systems can demonstrate procedural regularity — decisions made under an announced rule set consistently applied (Roehl & Hansen, 2024; Kroll et al., 2017).

Against the four SAC modalities, deterministic systems show minimal vulnerability: evidence gaps are unlikely (logic fully specified), responsibility diffusion is contained (authorship and deployment discretely attributable), decision opacity is absent by definition, and feedback failure remains possible but organizational rather than architectural. The DES natively supports this architecture through *logic_type=rule_based*; *override_occurred* is satisfied by construction, but cryptographic *hash_chain* and *sequence_number* require supplementary logging infrastructure not produced by rule logic alone, which is why *temporal_metadata* rates *partially_fillable*. Deterministic systems still serve as the baseline: five of six DES properties populated by construction, the highest ratio in Table 2.

### 3.2. Classical ML Systems with Human Oversight

Classical ML systems — versioned supervised ML inference with stable feature reconstruction, deployed with human-in-the-loop oversight — introduce the first governance complications. This scope excludes online-learning regimes, stochastic preprocessing, and feature-store time-travel gaps, which weaken *decision_logic* recoverability and require case-by-case assessment. The learned decision boundary is not readable as a rule set, creating a structural opacity that deterministic systems lack; deep learning in particular is often described as hiding internal logic from users (Busuioc, 2021). This opacity does not prevent evidence collection but changes what evidence must contain: the trace must capture model version, feature-pipeline version, $pre/post-processing$ transformations, and the confidence score alongside the inputs. DES v0.3.0 supports this through its *model_inference* object within *decision_logic* (*model_id*, *model_version*, *feature_vector_hash*, *confidence*) and the *logic_type=ml_inference* classifier.

Human-in-the-loop design inserts judgment at defined points to control risk, preserve contestability, and maintain accountability (Daruna, 2026). Each override creates an attributable DES event, but the governance value depends on whether it is substantive or ceremonial — the framework cannot distinguish these from trace data alone.

Against the SAC modalities, classical ML systems show moderate vulnerability. Evidence gaps emerge when feature engineering is not fully logged or when model training data provenance is incomplete. Responsibility diffusion increases because the decision involves both the model (trained by data scientists) and the reviewer (operating under time pressure).

Decision opacity is the primary new challenge — the model's learned boundary cannot be explained by citing rules, requiring additional explainability artifacts. Feedback failure risk increases because model performance degrades silently through data drift without triggering rule-level alerts.

The N4 framework accommodates classical ML systems through the DES's native *decision_logic* property and through label-free monitoring, which detects the distribution shifts that precede model degradation.

## 3.3. Hybrid ML+Rules Systems

Hybrid systems — the architecture most common in production risk decision systems — combine learned models with deterministic business rules at decision boundaries (e.g., a fraud pipeline scoring transaction risk and applying rules to choose *block/review/approve* based on score plus contextual factors). This creates governance seams: points where trace continuity and responsibility attribution bridge two paradigms — opaque ML scoring meets transparent rule logic. Decision boundaries that separate thresholds from core processing allow adjustment without changing the pipeline and maintain governance through documented decision criteria (Kasi, 2025); adequacy at the seam depends on whether the trace captures enough context to evaluate the score-rule interaction.

Consider a concrete failure: a transaction scores 0.72, below the 0.75 review threshold, so it passes. If later proven fraudulent, reconstruction must ask whether the model scored correctly, whether the threshold was appropriate for this transaction type, and whether the score-threshold interaction created an evidence gap. The DES must therefore capture score, threshold, rule evaluation path, and contextual factors informing the threshold.

Against the SAC modalities, hybrid systems exhibit compound vulnerability. Evidence gaps occur at the seam where model and rule traces are logged by different subsystems with different schemas and retention. Responsibility diffusion compounds because model performance is owned by data science while rule configuration is owned by business or compliance, and neither may fully understand the governance implications of the interaction. Decision opacity is partial (rules transparent, score not). Feedback failure is architecturally amplified: outcome labels may take weeks to arrive, during which both components continue operating on potentially degraded evidence.

The N4 framework addresses hybrid systems through seam-aware trace design that explicitly captures the handoff between ML and rules components. This is the architecture type for which the framework was originally designed, and it provides the most thorough governance coverage short of deterministic systems.

## 3.4. Agentic AI Systems

Agentic AI systems represent a qualitative departure. As used here, "agentic AI" refers specifically to multi-agent systems in which an orchestrator decomposes goals into sub-tasks, delegates them to specialized sub-agents (which may further delegate), invokes external tools, and synthesizes results through multi-step reasoning chains using foundation models with stochastic decoding — encompassing compound AI systems, LLM-based agent architectures, and autonomous multi-agent workflows. It excludes simpler single-agent systems with deterministic tool use and tightly constrained agents with fixed delegation patterns and pre-specified action spaces; such systems may exhibit some but not all of the breaks identified below. The zero-fillable profile in Table 2 applies to the multi-agent, foundation-model-based subclass defined here.

Where deterministic, ML, and hybrid systems share a bounded decision event that can be captured in a trace, agentic systems fragment decisions across multiple autonomous agents with dynamic tool use, delegated sub-tasks, and emergent coordination. Traditional governance models were built for human-operated systems with clear command hierarchies and discrete decision points, rendering them misaligned with continuous, distributed, opaque decision-making (Barot, 2025): the agentic decision is the emergent product of the entire delegation chain, with no single identifiable point for the DES to capture.

Kroll et al.'s insight about ML — that learned rules raise problems because they lack designer-fixed decision rules (Kroll et al., 2017) — applies with greater force to agentic systems. Where ML at least provides a fixed, versionable model, agentic reasoning is non-deterministic: prompt variation, reflection, delegation, and stochastic decoding routinely produce different outputs for unchanged semantic intent (Fatmi, 2026), undermining reconstructability architecturally.

Against the SAC modalities, agentic systems show severe vulnerability across all four. Evidence gaps are structural: execution logs record effects (API calls, state mutations) but do not capture the authorization rationale that allowed each event to occur (Fatmi, 2026). Responsibility diffusion reaches a qualitative threshold: the agentic AI has its own operational autonomy, creating an accountability vacuum in which responsibility evaporates, and the unstructured nature of its tasks, coupled with broad autonomy and inherently opaque training processes, defies traditional concepts of liability (Mukherjee & Chang, 2025). Decision opacity compounds because each agent in the chain may use different models, tools, and reasoning strategies. Feedback failure is amplified by the difficulty of attributing outcomes to specific agents or decisions within the chain.

The N4 framework identifies these gaps but does not yet provide remediation. The DES assumes a bounded decision event; agentic systems require a trace protocol that captures delegation graphs. Sufficiency metrics assume a single decision context; agentic systems distribute context across agent-local states. Label-free monitoring assumes stable decision distributions; agentic systems exhibit inherent variability that is difficult to distinguish from degradation. Section 5 examines these structural breaks in detail and proposes extensions to address them.

### 3.5. Summary: The Governance Coverage Gradient

Table 2 formalizes the governance coverage gradient using a purpose-built governance evaluation framework (Solozobov, 2026d). Each architecture-property pair is rated against the six DES v0.3.0 property groups using four categories:

- **fillable**: populated from artifacts the architecture produces by construction, without additional instrumentation.
- **partially_fillable**: recoverable only through supplementary artifacts (explainability tools, additional logging, manual enrichment) that the architecture does not produce by default.
- **unfillable**: structurally unrecoverable given the architecture's design constraints, regardless of instrumentation.
- **opaque**: the evidence exists internally but cannot be externalized — the behaviour is produced without any artifact exposing the generating state.

Borderline cases are resolved by requiring that structural properties guarantee artifact availability for the higher category; instrument-dependent evidence is rated *partially_fillable* regardless of how common such instrumentation is in practice. The fillable ratio is the proportion of the six properties rated fillable. The *override_escalation_record* property generalizes the original DES *human_override_record* (Solozobov, 2026a) to any structured authority intervention. DES v0.3.0 makes *override_occurred* a mandatory boolean across all tiers and logic types, so override tracking no longer depends on organizational discipline. Hybrid ML+Rules rates fillable because the rules component emits structured escalation context (triggering rule, threshold exceeded, risk category); Classical ML + HITL rates *partially_fillable* because override triggers remain discretionary even when *override_occurred* is populated. The distinction is structured versus unstructured escalation context, not substantive versus ceremonial cognition.

A worked contrast illustrates the coding scheme. For *decision_logic* under Classical ML + HITL, the logic IS the trained weights plus the inference pipeline — versioned, stored, and deterministically reproducible from the same input, therefore fillable. For *decision_boundary* under the same architecture, explainability tools provide approximations (SHAP values, feature importance) rather than direct readings of the learned surface, therefore *partially_fillable*. The same logic applies to all 24 cells; Table 1 records the complete coding rationale, and extended discussion is documented in the governance evaluation framework repository (Solozobov, 2026d).

**Table 1. Complete coding rationale (24 cells).** Each cell shows the rating and the structural reason. Abbreviated property names map to Table 2: context = *decision_context*, logic = *decision_logic*, boundary = *decision_boundary*, quality = *decision_quality_indicators*, override = *override_escalation_record*, temporal = *temporal_metadata*.

| Property | Det. Rules | Classical ML + HITL | Hybrid ML+Rules | Agentic AI |
|---|---|---|---|---|
| context | **fillable** — rules define context by construction | **partially** — input provenance needs supplementary logging | **partially** — rules context explicit; ML feature provenance needs instrumentation | **partially** — fragmented across agent-local contexts |
| logic | **fillable** — rules ARE the logic; versioned and readable | **fillable** — trained weights + inference pipeline are versioned and deterministically reproducible | **fillable** — rule logic and model weights both versioned | **opaque** — foundation-model reasoning is stochastic and internally inaccessible |
| boundary | **fillable** — explicit thresholds in rule definitions | **partially** — SHAP / feature-importance approximate the learned boundary | **fillable** — rule thresholds at decision seams are auditable | **unfillable** — decision surfaces emerge across multi-step reasoning; no boundary artifact |
| quality | **fillable** — rule outputs are exact pass/fail | **partially** — confidence scores are uncalibrated by default | **partially** — exact signals from rules; partial from ML | **opaque** — per-agent signals internal; cross-agent aggregation undefined |
| override | **fillable** — structured escalation; *override_occurred* mandatory in DES v0.3.0 | **partially** — *override_occurred* populated but review criteria discretionary | **fillable** — structured rule-triggered escalation with threshold, risk category, authority | **unfillable** — no mandate boundaries across delegation chains |

| Property | Det. Rules | Classical ML + HITL | Hybrid ML+Rules | Agentic AI |
|---|---|---|---|---|
| temporal | **partially** — deterministic single timeline is architectural, but *hash_chain* and *sequence_number* require cryptographic logging instrumentation, not produced by rule logic alone | **partially** — timestamps exist but *hash_chain* needs instrumentation | **partially** — ordering clear within components; seam temporal coordination lacking | **partially** — per-agent timestamps local; cross-agent ordering uncoordinated |

**Table 2. Cross-architecture governance evidence feasibility (governance evaluation framework)**

| DES Property | Deterministic Rules | Classical ML + HITL | Hybrid ML+Rules | Agentic AI |
|---|---|---|---|---|
| *decision_context* | fillable | partially | partially | partially |
| *decision_logic* | fillable | fillable | fillable | **opaque** |
| *decision_boundary* | fillable | partially | fillable | **unfillable** |
| *decision_quality_indicators* | fillable | partially | partially | **opaque** |
| *override_escalation_record* | fillable | partially | fillable | **unfillable** |
| *temporal_metadata* | partially | partially | partially | partially |
| **Fillable ratio** | **0.83** | **0.17** | **0.50** | **0.00** |
| **Opaque ratio** | **0.00** | **0.00** | **0.00** | **0.33** |

An interpretive caveat: the fillable ratio is a conservative proxy for by-construction evidence availability, not a direct measure of governance adequacy. Organizational failures, adversarial drift, and counterfactual unavailability (Section 6) can undermine governance regardless of how completely the architecture produces evidence artifacts. The ratio measures the architectural starting point, not the governance outcome.

A methodological note on the fillable ratio: equal weighting across the six properties reflects that each captures a distinct governance requirement whose absence creates a specific accountability blind spot. The binary threshold (fillable vs. not) is conservative by design — it credits only evidence produced by construction. An ordinal alternative (*fillable=1.0*, *partially_fillable=0.5*, *unfillable=0*, *opaque=0*) yields 0.92/0.75/0.58/0.17 for *Deterministic/Hybrid/Classical ML/Agentic* respectively; the architecture ordering is preserved and the gradient remains monotonic, confirming P1's robustness to reasonable alternative weightings.

The gradient reflects architectural governance capacity, not a simple autonomy axis (Solozobov, 2026b; Solozobov, 2026a; Solozobov, 2026c; Solozobov, 2026e). Two qualitative features accompany the ordering: opaque-rated DES cells appear only in agentic systems in Table 2 (Classical ML is structurally opaque as a property, but the evidence-availability rating distinguishes "opaque" from "partially fillable" based on whether supplementary instrumentation can externalize the behavior), and the hybrid-to-agentic transition introduces unfillable and opaque ratings that represent structural impossibilities rather than resource constraints. These fillability assessments measure by-construction evidence production; Section 6 identifies boundary conditions (counterfactual unavailability, adversarial drift, the opacity–determinism spectrum) that limit governance effectiveness regardless of fillability scores. Section 4 examines how the unfillable and opaque entries in Table 2 arise dynamically through cascading failures across framework layers.

This gradient yields the following analytical proposition:

P1: Governance evidence feasibility follows the ordering deterministic rules (0.83) > hybrid ML+rules (0.50) > classical ML + HITL (0.17) > agentic AI (0.00). Notably, this ordering is non-monotonic with respect to system autonomy: hybrid systems are more automated than classical ML + HITL (no mandatory human review), yet score higher on governance coverage because the deterministic rules component produces governance artifacts by construction — explicit decision boundaries, structured escalation triggers, and version-controlled rule logic — that pure ML systems do not. The gradient therefore reflects architectural governance capacity (how many DES properties the architecture fills by construction) rather than a simple autonomy axis. The gradient collapses entirely at the agentic boundary, where no DES property is fillable from available governance artifacts.

The ordering is robust to plausible alternative codings. The deterministic-first position is invariant to any single-cell change; the agentic-last position is invariant up to possible ties (a single agentic cell upgraded to fillable would move agentic to 0.17 and tie with classical ML, not surpass it). Inverting the hybrid-over-classical gap would require downgrading at least three hybrid cells or equivalently recoding multiple classical cells. The gradient is therefore an ordinal tendency robust to reasonable alternative judgments, not an artifact of borderline coding decisions.

# 4. The Cascade of Uncertainty

The cross-architecture comparison in Section 3 treated each governance challenge in isolation — the unfillable and opaque entries in Table 2 identify which DES properties each architecture cannot populate. In practice, however, these challenges do not arise independently. This section introduces the cascade of uncertainty: a structural property of decision systems in which governance failures at one layer propagate downward, amplifying evidence gaps at subsequent layers. The cascade connects the four N4 framework components into a dependency chain where upstream failures compound into downstream governance collapse.

## 4.1. The Cascade Mechanism

The cascade operates through five linked stages, each of which degrades a different governance property:

**Stage 1: Feature Incorrectness.** The decision pipeline receives input data that is incorrect, stale, or incomplete. In a fraud detection system, this might mean that a customer's transaction history is missing recent activity because of a data pipeline delay, or that a risk feature is computed from a table that has not been refreshed. The decision system processes these inputs without awareness of their deficiency — the features arrive in the expected format and pass schema validation, but their content does not reflect the actual state of the world.

**Stage 2: Invisible Errors.** Incorrect features produce incorrect decisions. In risk-scoring domains where data degradation manifests as missing or stale risk-elevating signals, the directional bias is toward false negatives — cases that should have been flagged but were not (a transaction that would score 0.85 risk with correct features scoring 0.62 with stale data, falling below the review threshold). False positives typically generate downstream governance artifacts (human reviews, appeals, case files) that make them traceable — but this traceability is conditional, not architectural: in domains where affected subjects lack accessible appeal pathways (automated welfare suspensions, gig-worker deactivations, preventive-policing flags on marginalized populations), false positives also fail to produce downstream artifacts and cascade into the same epistemic indistinguishability as false negatives. The governance-invisible path is therefore the broader class of decisions for which no corrective feedback loop reaches the governance layer, for reasons architectural (false negatives) or sociotechnical (unappealed false positives).

**Stage 3: Epistemically Indistinguishable Decisions.** True positives generate case files, review records, and outcome documentation; false negatives produce no downstream artifacts. The decision trace exists but is indistinguishable from a correctly scored low-risk transaction. The governance archive is technically complete but epistemically empty.

**Stage 4: Undetectable Degradation.** As false negatives accumulate, score distributions, decision volumes, and override rates may remain stable. Label-free monitoring (Solozobov, 2026e) detects distribution shifts in features and scores, but degradation that operates through data quality rather than model drift may not trigger the monitoring signal.

**Stage 5: Cumulative Losses.** Undetected false negatives accumulate into material losses until an external signal (a complaint, a regulatory audit, a cross-system pattern) reveals the gap. Identifying which archived decisions were affected by feature incorrectness requires two evidential channels: ground truth outcome labels and independent data provenance or source-system reconciliation. The former may never arrive; the latter requires cross-system infrastructure that most pipelines do not maintain.

The cascade is not hypothetical: organizations constructing complete decision lineages frequently discover that this requires matching events across systems with dissimilar schemas, time granularities, and retention policies (Bisht, 2026) — a direct consequence of cascade-stage evidence loss.

## 4.2. Mapping the Cascade to the N4 Framework

Each cascade stage corresponds to a specific component of the N4 framework, revealing that the framework's four layers are not independent defences but a serial dependency chain:

**Stage 1 (Feature Incorrectness) maps to the decision trace schema** (Solozobov, 2026a). The DES captures input features at decision time, creating a record that could, in principle, be used to detect feature incorrectness retrospectively. However, the DES records feature values as received by the decision system — it does not independently verify feature correctness against source systems. This is an inherent limitation: the trace is faithful to what the system saw, not to what the system should have seen. Evidence-driven governance systems require that documentation extend beyond high-level descriptions to include versioned configurations, feature selection rationales, and known limitations, supporting reproducibility under comparable conditions (Nwaodike, 2022).

**Stage 2 (Invisible Errors: False Negatives and Unappealed False Positives) maps to the evidence sufficiency model** (Solozobov, 2026c). Sufficiency metrics evaluate whether the governance evidence archive contains enough information to support accountability judgments. The sufficiency question takes two forms in this stage: for false negatives, whether the archive can distinguish a correctly scored low-risk transaction from an incorrectly scored one; for unappealed false positives, whether it can distinguish a correctly scored high-risk case from an incorrectly scored but never-contested one. In the absence of ground truth labels or external contestation signals, both distinctions may be unresolvable —

the sufficiency metrics indicate the archive is complete, but completeness does not guarantee correctness.

**Stage 3 (Epistemic Indistinguishability) maps primarily to the decision trace schema and evidence sufficiency model** (Solozobov, 2026a; Solozobov, 2026c). The trace records structurally valid but semantically incorrect decisions — the evidence exists but cannot distinguish correct from incorrect outcomes. The root problem is epistemic: the DES faithfully records what the system saw, and the sufficiency metrics report the archive as complete, but completeness does not guarantee correctness. The governance evidence archive is technically adequate yet epistemically empty. The label-free monitoring layer (Solozobov, 2026e) provides a secondary defence at this stage, but as discussed below, its detection capability depends on the degradation channel.

**Stage 4 (Undetectable Degradation) maps to the label-free monitoring layer** (Solozobov, 2026e). This is the stage where the monitoring layer should detect the accumulating governance failure. Standard distribution-shift monitoring does catch many pipeline failures: stale data tables compress variance toward zero, and missing activity spikes zero-count features, both producing detectable distribution shifts. The cascade's blind spot is narrower but more dangerous: distribution-preserving corruption, where semantic errors maintain identical statistical properties to correct data. Adversarial drift, systematic data-entry errors mirroring historical patterns, or upstream logic changes preserving marginal distributions while altering conditional relationships — these create governance failures that pass through marginal distribution monitors undetected (multivariate joint-distribution detectors may still catch them, but production monitoring stacks default to per-feature univariate tests). This suggests that label-free monitoring should incorporate semantic data quality signals alongside statistical distribution signals — a framework extension identified as future work in the label-free monitoring study (Solozobov, 2026e).

**Stage 5 (Cumulative Loss) maps to the governance evidence framework** (Solozobov, 2026b). The cascade's terminal stage is precisely what the SAC diagnosis describes: a condition in which architectural properties exceed the design capacity of governance infrastructure. The cascade demonstrates that SAC is not merely a static vulnerability but a dynamic process — governance evidence degrades progressively through linked stages, each building on the failures of the preceding one.

The cascade mapping reveals a critical insight: the N4 framework's four components are necessary but may not be jointly sufficient. Each component addresses its own stage effectively in isolation, but the cascade creates failure modes that emerge from the interaction between stages. A trace that faithfully records incorrect features, a sufficiency metric that reports completeness for epistemically empty archives, monitoring that does not fire because the degradation channel differs from the detection channel — these are not individual failures but systemic ones that arise from the cascade structure itself.

The cascade mechanism yields a testable proposition:

P2: In risk-scoring decision pipelines where governance evidence flows through sequential processing stages (feature engineering, model inference, post-processing, human review), an evidence gap at stage k produces compounding governance degradation at downstream stages. The severity of the evidence deficit increases at each stage because downstream governance artifacts inherit the epistemic limitations of upstream records — a decision trace built on incorrectly recorded features is governance-deficient regardless of how completely the downstream stages log their own processing. This cascade is demonstrated here for the false-negative path in binary risk-scoring systems; whether analogous compounding occurs in other pipeline topologies (e.g., false-positive paths, multi-output classification, or non-sequential architectures) requires separate investigation. Empirical measurement of the compounding effect would require external ground-truth injection at each stage to distinguish inherited from locally introduced evidence gaps.

### 4.3. Implications for Framework Design

The cascade has two practical implications. First, governance framework design must account for cross-layer dependencies: testing each N4 component in isolation may report success while the cascade creates compound failures that no single component detects. Current audit practices focus on individual model components rather than cross-layer evidence chains (Mökander et al., 2021). Second, the cascade intensifies with architectural complexity but no architecture is immune to its early stages: deterministic systems still suffer from factually wrong inputs even when rule logic is transparent; in agentic systems, autonomous feature gathering and multi-step reasoning multiply the stages at which incorrectness enters and propagates. The next section examines how these compounding effects create structural breaks in agentic AI.

## 5. Agentic Extension

Section 3 identified agentic AI as the only architecture with zero fillable DES properties, and Section 4 showed how governance failures compound across framework layers. This section examines the specific structural breaks agentic systems introduce and whether the N4 framework can be extended to address them. Agentic systems differ from prior architectures in three structural properties — decision diffusion, evidence fragmentation, and responsibility ambiguity — each breaking a foundational assumption of the N4 chain.

### 5.1. Decision Diffusion

The N4 DES assumes a bounded decision event: inputs processed, model or rule set applied, output produced. This holds for deterministic, ML, and hybrid systems; in agentic systems it fails. An orchestrator receives a goal, decomposes it into sub-tasks, delegates them to specialized agents, reasons about intermediate results, potentially re-delegates, and produces a composite output. The decision is a directed acyclic graph (DAG) of delegated sub-decisions when the execution trace is unrolled over time, each node with its own inputs, reasoning, and outputs. The underlying control flow may be cyclic (ReAct loops, reflection, state-machine delegation); "DAG" here refers to the execution trace exposed to the governance layer, not the agent's internal policy graph. Accountability is distributed across multiple stakeholders, each with partial influence (Tatipamula, 2025).

This fragmentation instantiates the many hands problem (Nissenbaum, 1996) — in AI, the impossibility of pinpointing individual responsibility when multiple actors and processes jointly produce outcomes (Novelli et al., 2024). A single trace record cannot capture a DAG; agentic systems require a protocol capturing delegation, rationale, and composition at each step, since without execution-time boundaries provenance is incomplete and downstream effects ungovernable (Fatmi, 2026).

### 5.2. Evidence Fragmentation

Even if a trace protocol captured the DAG, a second structural break prevents assembly: evidence fragmentation. In non-agentic systems, governance artifacts share a common decision event identity and can be assembled through schema mapping even when subsystems differ (Section 3.3). In agentic systems, each agent maintains its own local context and logs, with ephemeral and heterogeneous schemas lacking a shared decision identity; evidence for a single composite decision is scattered across agent-local contexts, tool invocation logs, and inter-agent messages.

As established in Section 3.4, agent reasoning is non-deterministic and execution logs capture effects rather than authorization rationale. An execution log answers what happened; governance evidence must answer why the action was authorized. In agentic systems the authorization rationale lives in ephemeral agent prompts, context windows, and reasoning chains discarded after task completion. Observability must therefore extend beyond logging to include monitoring of decision chains, action sequences, and environmental feedback loops (Joshi, 2025).

The evidence sufficiency model (Solozobov, 2026c) assumes completeness can be evaluated against a known schema. In agentic systems the schema is dynamic — different task decompositions produce different DAGs with different evidence requirements — so sufficiency metrics must evaluate not only field population but DAG completeness and preservation of causal links between agent-local fragments.

## 5.3. Responsibility Ambiguity

The third structural break concerns attribution. SAC (Solozobov, 2026b) treats responsibility diffusion as one of four bounded modalities; in agentic systems it becomes the dominant failure mode.

In human organizations, hierarchical authority mitigates the many hands problem: the chain of command determines who bears ultimate responsibility. Multi-agent architectures lack this mitigation — control is distributed across autonomous agents, none individually determining the outcome (Joshi, 2025). An orchestrator has nominal authority but limited visibility into sub-agent reasoning; the sub-agent has operational control but no broader context; neither bears clear responsibility. At the human-system boundary, such vacuums can create moral crumple zones where human operators absorb liability for failures they could not prevent (Elish, 2019), but within the agent system the problem is structural: no attribution mechanism exists.

In multi-agent systems, the organizational many-hands problem (Ada Lovelace Institute; AI Now Institute; Open Government Partnership, 2021) compounds with the agent-level one: responsibility must be traced across human designers, operators, orchestrators, and sub-agents. Remediation requires a delegated attribution model specifying how evidence of delegated contribution flows through delegation chains.

## 5.4. Toward Framework Extensions

The three breaks are not independent: they interact through the cascade of Section 4 — diffusion creates multiple entry points for feature incorrectness, fragmentation prevents assembling a complete monitoring picture, and ambiguity prevents post-incident accountability assignment even when evidence is eventually gathered. Addressing them requires three corresponding extensions to the N4 framework:

**Inter-agent decision trace protocols.** The DES must be extended from a single-event schema to a delegation-aware protocol that captures the full delegation DAG. Each agent in the chain must emit trace records including the task received, a structured decision rationale (policy consulted, constraints evaluated, tool or sub-agent selection justification), the sub-tasks delegated, and the output produced; correlation identifiers must link parent and child traces. The framework does not require disclosure of raw private chain-of-thought: governance-relevant rationale is a structured record of what mandate the agent was operating under and which constraints it checked, not a transcript of the model's internal reasoning tokens. DES v0.3.0 provides partial support through its *decision_boundary* property with *upstream_decisions* and *downstream_consumers* arrays and per-interface *boundary_contract* objects — the agentic extension requires generalizing these from static contracts to dynamic delegation graphs whose coupling is determined at runtime. Correlation-based distributed tracing is a well-established pattern (Sigelman et al., 2010), but governance traces must additionally capture semantic authorization rationale — why the agent was mandated to act and what constraints it evaluated — which is what transforms an execution trace into governance evidence (Joshi, 2025).

**Distributed evidence aggregation.** Sufficiency metrics must be extended to evaluate evidence completeness across a delegation graph rather than within a single decision context. This requires a mechanism for assembling agent-local evidence fragments into a unified governance record, preserving causal links between fragments, and detecting gaps where agent-local evidence was not captured or was discarded. The aggregation mechanism must handle the heterogeneity of agent evidence — different agents may use different models, tools, and data formats, producing evidence in incommensurable schemas.

**Delegated responsibility attribution.** The responsibility diffusion modality must be extended to capture the evidence required for accountability judgments in delegation chains. Specifically, governance traces must record: (a) the delegation parameters (what task was delegated, with what constraints), (b) mandate boundary conditions (what the sub-agent was authorized to do), and (c) outcome provenance (which agent produced which component of the composite result). The framework is calibrated to mandate-based, fault-based, and principal-agent accountability structures; strict-liability regimes that require only causal execution and outcome evidence do not need the rationale and mandate fields specified here. Within these structures, whether a principal-agent attribution model or an alternative is adopted remains a normative question beyond the framework's scope — the framework's contribution is specifying the evidence required to support those attribution judgments, ensuring accountability remains determinable rather than structurally unresolvable.

These three extensions address decision diffusion, evidence fragmentation, and responsibility ambiguity respectively. A fourth SAC modality — feedback failure — is amplified in agentic systems (Section 3.4) but is partially addressed by the extensions: the delay aspect remains structural (outcome signals are delayed or absent regardless of how many agents produced the decision), while the attribution aspect becomes evidentially traceable under delegated responsibility attribution, which records the provenance chain linking outcomes to agent actions. The framework supplies evidence for attribution judgments; it does not itself assign responsibility. The distributional monitoring approach (Solozobov, 2026e) applies once inter-agent traces provide the evidence substrate, though Section 4 identifies a blind spot — data-quality degradation that does not manifest as distribution shifts — requiring data-quality signals regardless of architecture.

These extensions are proposed as analytical constructs, not implemented artifacts; their feasibility and practical implications require empirical investigation and constitute a primary direction for future work. The structural breaks and proposed extensions yield two analytical expectations that follow from the framework's structure:

P3: Responsibility attribution in multi-agent systems behaves in two regimes relative to inter-agent trace protocols. **Given** protocols, attribution is possible but its complexity scales monotonically with DAG depth and branching — the number of agents whose contributions must be disentangled grows with delegation depth and branching factor, at a rate depending on DAG topology (subgraph reuse, hierarchical vs cross-agent synthesis). **Absent** protocols, attribution is not merely complex but structurally impossible (§5.2 accountability vacuum) regardless of graph size. The analytical expectation is twofold: protocol-enabled attribution scales with topology; protocol-absent attribution fails categorically.

P4: Within a cooperative agent ecology — where all agents are white-box or governance-trace-compatible (see Section 6.4 for the proprietary-API boundary) — the three extensions would improve four of six DES properties:

- *decision_context* (*partially_fillable*, strengthened): inter-agent traces add delegation parameters and structured rationale records. Like *decision_quality_indicators*, self-reported rationale is subject to the unfaithful-explanation problem — foundation models can produce post-hoc rationalizations citing constraints that had no causal bearing on the output — so to function as governance evidence rather than unverified agent output, rationale records must be anchored by deterministic execution proofs (API constraint checks, structured-output gates, state-machine transitions) rather than relied upon as self-standing traces;
- *temporal_metadata* (*partially_fillable*, strengthened): distributed evidence aggregation reconstructs ordered cross-agent timelines; satisfying DES v0.3.0's *hash_chain* and *sequence_number* integrity across the graph is a non-trivial distributed-consensus problem;
- *override_escalation_record* (unfillable -> *partially_fillable*): delegated responsibility attribution supplies structured mandate boundaries, verifiable to the extent agents expose constraint-checking logic;
- *decision_quality_indicators* (opaque -> *partially_fillable*): agents emit self-reported quality signals (confidence, uncertainty, calibration) as mandatory trace fields. As with *decision_context*, self-report is necessary but not sufficient: it must be anchored by externally auditable calibration metadata, evaluation provenance, or cross-agent consistency checks; otherwise it is closer to an execution log than to governance evidence.

All four reach *partially_fillable* rather than fillable because the extensions are supplementary instrumentation (per §3.5's taxonomy). *decision_logic* (opaque) and *decision_boundary* (unfillable) remain structurally limited in foundation-model-based agents. The post-extension profile (4 partially, 1 opaque, 1 unfillable) approaches but does not equal the hybrid ML+rules baseline (3 fillable, 3 partially).

# 6. Boundary Conditions and Limitations

The preceding sections showed that the N4 framework, originally formulated for hybrid ML+rules systems, can be analytically transferred to deterministic and classical ML architectures while encountering structural breaks in the agentic case. This section defines four boundary conditions where the framework's governance coverage is structurally limited. These boundaries are not implementation gaps to be filled but inherent constraints that any post-incident governance framework must acknowledge.

## 6.1. The Opacity–Determinism Spectrum

The framework's governance coverage is constrained by two independent architectural properties: structural opacity (whether decision logic can be inspected) and non-determinism (whether the decision can be reproduced). Classical ML inference is deterministic but opaque, which is why it scores lower (0.17) than hybrid systems (0.50) despite identical determinism. Deterministic rule engines are both transparent and deterministic; foundation models with stochastic decoding are both opaque and non-deterministic (Fatmi, 2026), and agentic systems compound both through multi-step delegation chains.

Opacity limits evaluability: auditors cannot assess whether decision logic was appropriate without supplementary explainability artifacts. Non-determinism limits reconstructability: the DES records what happened on one occasion but cannot guarantee repeatable output. For hybrid systems the rules component compensates for ML opacity at governance seams; for agentic systems, where both opacity and non-determinism are architectural rather than parametric, even weak reconstructability requires the inter-agent trace protocols proposed in Section 5.

## 6.2. Counterfactual Unavailability

The evidence sufficiency model (Solozobov, 2026c) assumes that correct accountability judgments are possible in principle — that ground truth labels will eventually arrive or that proxy signals can substitute. In some domains, this assumption fails permanently.

Hiring decisions, welfare benefit determinations, and preventive policing share a common property: the counterfactual outcome is structurally unobservable, a problem recognized in algorithmic fairness as the selective labels problem where observed outcomes exist only for cases past the decision gate (Schwartz et al., 2022). Sufficiency metrics can confirm that the archive is structurally complete (all DES fields populated) but not that the decision was adequate — sufficiency measures completeness, not correctness. Label-free monitoring (Solozobov, 2026e) detects degradation relative to a baseline but cannot establish whether the baseline itself was adequate.

## 6.3. Adversarial Drift

Label-free monitoring assumes that degradation manifests in detectable distributional changes. A sophisticated adversary can preserve the statistical properties of governance artifacts while degrading actual decision quality — narrow audits that miss such exploitation amount to gaming and provide false assurance rather than genuine accountability (Costanza-Chock et al., 2022). In fraud detection, an adversary who understands the monitoring signals can structure fraudulent transactions to preserve score distributions, override rates, and feature statistics while the underlying fraud rate increases: the archive remains statistically normal, the monitoring layer reports no anomalies, yet the system is systematically failing.

The framework does not claim adversarial robustness. Its contribution is diagnostic: identifying when governance evidence has degraded, not defending against deliberate manipulation. Adversarial robustness requires additional mechanisms (anomaly detection on governance metadata, cross-system correlation, external audit) outside the framework's current scope.

### 6.4. Proprietary API Boundaries

The inter-agent trace protocol of Section 5 assumes every agent can be instrumented to emit structured traces. In production multi-agent architectures, sub-agents are frequently third-party commercial LLM endpoints that are black-boxed by the vendor. The orchestrator cannot compel a proprietary endpoint to externalize reasoning traces, constraint-checking logic, or confidence calibration; when delegation crosses a commercial vendor boundary, governance opacity re-emerges regardless of orchestrator-side instrumentation. The inter-agent trace protocol therefore achieves its improvements only within cooperative or white-box agent ecologies — a condition current commercial ecosystems do not guarantee. Addressing this boundary requires vendor-side governance-trace standards or governance-preserving API wrappers that extract observable behavioral signals without internal state access.

### 6.5. Implications for Framework Scope

These four boundaries define the framework's valid operating envelope for strong accountability judgments: deterministic and hybrid ML+rules systems with bounded non-determinism, domains where ground truth eventually arrives, and environments where degradation is organic rather than adversarial. Outside this envelope a weaker diagnostic use remains available — the framework can still identify what evidence exists and what is missing, even in counterfactual-unavailable or adversarial domains — but evidence completeness no longer underwrites definitive accountability decisions. Specifying these limits is itself a contribution: a framework that delineates where it applies provides more guidance than one claiming universal applicability.

## 7. Discussion and Future Work

### 7.1. Research Contributions

This paper completes a research program proceeding from diagnosis through engineering to measurement and synthesis, and builds on it with three additional contributions: analytical testing of the framework's transferability across architectures, identification of compound failure modes through the cascade of uncertainty, and mapping of the structural boundaries that emerge when the framework confronts agentic AI (Solozobov, 2026b; Solozobov, 2026a; Solozobov, 2026c; Solozobov, 2026e). The integrated contribution is a governance evidence infrastructure — from diagnostic theory to operational measurement — analytically developed for hybrid ML+rules systems and extended to agentic architectures. Empirical validation remains future work; the present contribution establishes the analytical foundations and identifies the structural constraints that shape possible governance solutions.

### 7.2. Relationship to the Accountability Literature

The framework is complementary to, not a replacement for, the broader algorithmic accountability literature. Normative scholarship — on the many hands problem, moral crumple zones, distributed responsibility, and sociotechnical accountability regimes — asks who should be accountable and under what conditions. The N4 framework asks a companion question: what evidence must be available to support accountability judgments, however they are assigned. Evidence-driven governance provides the operational substrate (Nwaodike, 2022) that normative judgments presuppose: without reconstructable decision traces, Nissenbaum-style attribution, Elish-style apportionment of liability, or Novelli-style contestation become practically underdetermined. The framework does not decide who is accountable; it specifies the evidence without which that decision cannot be grounded.

The framework is also post-incident rather than preventive, addressing the reconstruction problem — can the process be reconstructed from available evidence after an adverse outcome — and thereby providing the evidentiary infrastructure that traditional retrospective accountability (Eilstrup-Sangiovanni & Hofmann, 2024) requires. Prevention is a broader challenge beyond evidence architecture.

### 7.3. Regulatory Implications

The framework is potentially informative for emerging AI governance regulation. The EU AI Act requires high-risk AI systems to enable automatic recording of events over the system's lifetime and mandates that logging capabilities facilitate post-market monitoring (European Parliament and Council of the European Union, 2024). The N4 framework's DES provides a structured specification of what such logs must contain to support governance evidence requirements — going beyond generic logging mandates to specify the six decision event properties that enable decision reconstruction. For deterministic and hybrid systems, compliance with Article 12 logging requirements can be informed by DES field specifications, which provide a structured operationalization of what regulatory logging mandates require in practice. For agentic systems, however, the structural breaks identified in Section 5 reveal that Article 12 compliance faces the same challenges as the N4 framework itself: decision diffusion means no single log captures the complete decision, and evidence fragmentation means that the logs required for reconstruction are distributed across multiple agents. The inter-agent trace protocol proposed in Section 5 could potentially inform implementation approaches for Article 12 compliance in agentic deployments, though its formal specification remains future work.

### 7.4. Future Research Directions

The agentic extension proposed in Section 5 identifies three analytical constructs — inter-agent trace protocols, distributed evidence aggregation, and delegated responsibility attribution — that require empirical development. Three specific research directions follow.

First, formal specification of inter-agent trace protocols using process algebra or temporal logic formalisms. The delegation graph structure proposed in Section 5 needs formal properties: completeness (all delegation events are captured), consistency (parent and child traces agree on delegated tasks and results), and minimality (no redundant trace information that creates storage or privacy burdens). TLA+ or similar specification languages could express these properties and enable model checking against protocol implementations. Recent work on policy-as-code frameworks for continuous AI assurance demonstrates the feasibility of expressing governance obligations as executable, reproducible criteria with deterministic acceptance gates (Muhammad et al., 2026).

Second, empirical testing of the cascade of uncertainty using instrumented decision pipelines. The cascade model predicts specific failure propagation patterns — feature incorrectness producing false negatives producing untraceability producing undetectable degradation. These predictions are testable: deliberately introducing controlled data quality failures into an instrumented pipeline and measuring whether the N4 monitoring layer detects the cascade at each stage, and if so, how quickly.

Third, privacy-preserving governance evidence. The DES requires capturing detailed decision context, which may include personal data subject to GDPR, CCPA, or similar regulations.

A tension exists between governance evidence completeness (capture everything needed for reconstruction) and data minimization (capture only what is necessary for the specified purpose). Techniques such as differential privacy, zero-knowledge proofs, or purpose-limited access controls could reconcile these requirements, but their impact on governance evidence utility is unexplored.

### 7.5. Limitations of This Analysis

This paper is a theory-building contribution. The cross-architecture comparison is analytical, applying governance criteria to architecture types by structural properties rather than measured outcomes. The cascade of uncertainty generates testable predictions but has not itself been tested, and the agentic extensions are proposed as analytical constructs, not validated designs. These limitations are intentional: the contribution is identifying the right questions and the structural constraints that shape possible answers.

## 8. Conclusion

This paper synthesized the operational governance evidence framework developed across four prior studies (Solozobov, 2026b; Solozobov, 2026a; Solozobov, 2026c; Solozobov, 2026e) and analytically assessed its transferability across four decision system architectures. Three findings emerged.

First, the framework exhibits a governance coverage gradient. Deterministic rule-based systems satisfy five of six DES properties by construction (the sixth requires immutable-logging infrastructure common to all architectures). Classical ML systems remain governable only with supplementary instrumentation and explainability artifacts. Hybrid ML+rules systems — the framework's analytically developed domain — demand seam-aware governance design at the boundary between learned models and business rules. Agentic AI systems, however, expose structural breaks that the current framework cannot address without extension.

Second, governance failures propagate through a cascade of uncertainty. Feature incorrectness produces governance-invisible errors — false negatives and, in domains without accessible contestation, unappealed false positives — which generate epistemically indistinguishable decisions, causing undetectable degradation that accumulates into silent losses. The cascade maps the framework's four components — diagnostic theory, trace schema, sufficiency measurement, and label-free monitoring — into a serial dependency chain where upstream failures compound into downstream governance collapse. This cascade property means that the framework's components, while individually sound, may fail collectively in ways that no single component detects.

Third, agentic AI systems introduce three structural breaks: decision diffusion (no single decision point to trace), evidence fragmentation (governance artifacts scattered across agent-local contexts), and responsibility ambiguity (accountability cannot be attributed through delegation chains). These breaks require three corresponding framework extensions: inter-agent trace protocols, distributed evidence aggregation, and delegated responsibility attribution models. These extensions identify the structural requirements for agentic governance; the partial recovery of DES-property fillability they enable applies within cooperative, trace-compatible agent ecologies — where all agents can emit structured governance traces — rather than to all commercial deployments where proprietary API boundaries may re-introduce opacity. The extensions are proposed as analytical constructs requiring empirical development.

The gradient measures by-construction evidence availability, not end-to-end adequacy: the four boundary conditions define the limits beyond which architecturally complete evidence may be insufficient. Within these limits, the N4 chain provides a structured approach to the initiating question: when automated decision systems fail, what evidence is needed to reconstruct what happened and why?